\documentclass[preprint2]{aastex}

\newcommand{\Rsolar}{\mbox{$R_{\odot}\,$}}

\newcommand{\kms}{\mbox{$\mbox{km\,s}^{-1}$}\,}

\shorttitle{The angular size of the Cepheid $\ell$ Car}
\shortauthors{Kervella et al.}

\begin{document}

\title{The angular size of the Cepheid $\ell$ Car: a comparison
of the interferometric and surface brightness techniques}

\author{Pierre Kervella\altaffilmark{1,7},
Pascal Fouqu\'e\altaffilmark{1,2},
Jesper Storm\altaffilmark{3},
Wolfgang P. Gieren\altaffilmark{4},
David Bersier\altaffilmark{5},
Denis Mourard\altaffilmark{6},
Nicolas Nardetto\altaffilmark{6},
Vincent Coud\'e du Foresto\altaffilmark{7}
}

\altaffiltext{1}{European Southern Observatory, Casilla 19001,
Santiago 19, Chile}

\altaffiltext{2}{Observatoire Midi-Pyr\'en\'ees, UMR 5572, 14, avenue
Edouard Belin, F-31400 Toulouse, France}

\altaffiltext{3}{Astrophysikalisches Institut Potsdam, An der
Sternwarte 16, D-14482 Potsdam, Germany}

\altaffiltext{4}{Universidad de Concepci\'on, Departamento de
F\'{\i}sica, Casilla 160-C, Concepci\'on, Chile}

\altaffiltext{5}{Space Telescope Science Institute, 3700 San Martin
Drive, Baltimore, MD 21218, USA}

\altaffiltext{6}{GEMINI, UMR 6203, Observatoire de la C\^ote d'Azur,
Avenue Copernic, F-06130 Grasse, France}

\altaffiltext{7}{LESIA, Observatoire de Paris-Meudon, 5, place Jules
Janssen, F-92195 Meudon Cedex, France}

\email{pkervell@eso.org,
pfouque@eso.org,
jstorm@aip.de,
wgieren@coma.cfm.udec.cl,
bersier@stsci.edu,
Denis.Mourard@obs-azur.fr,
Nicolas.Nardetto@obs-azur.fr,
vincent.foresto@obspm.fr
}

\begin{abstract}
Recent interferometric observations of the brightest and angularly largest
classical Cepheid, $\ell$~Carinae, with ESO's VLT Interferometer (VLTI)
have resolved with high precision the variation of its angular diameter with phase.
We compare the measured angular diameter curve to
the one we derive by an application of the Baade-Wesselink type
infrared surface brightness technique, and find a
near-perfect agreement between the two curves.
The mean angular diameters of $\ell$~Car from the two techniques agree
very well within their total error bars (1.5\,\%), as do the derived distances (4\,\%).
This result is an indication that the calibration of the surface
brightness relations used in the distance determination of far away
Cepheids is not affected by large biases.
\end{abstract}


\section{Introduction}

Cepheid variables are fundamental objects for the calibration of the
extragalactic distance scale. Distances of Cepheids can be derived in
at least two different ways: by using their observed mean magnitudes
and periods together with a period-luminosity relation, or by applying
a Baade-Wesselink (hereafter BW) type technique to determine their distances
and mean diameters from their observed variations in magnitude, color
and radial velocity.  This latter technique has been dramatically
improved by the introduction of the near-infrared surface brightness
method (hereafter IRSB) by \citet{Welch94}, and later by
\citet{FG97} who calibrated the relation between the $V$-band surface
brightness and near-infrared colors of Cepheids. For this purpose, they used
the observed interferometric angular diameters of a number of giants and
supergiants bracketting the Cepheid color range. This method has
been applied to a large number of Galactic Cepheid variables, for instance
by \citet{GFG97}, \citet{GFG98}, and \citet{Storm04}.

Applying the surface-brightness relation derived from stable
stars to Cepheids implicitly assumes that the relation
also applies to pulsating stars.
The validity of this assumption can now be addressed by comparing
direct interferometric measurements of the angular diameter variation of
a Cepheid to the one derived from the IRSB technique. It has recently
been shown by \citet{Kervella04}, hereafter K04, that the VLT Interferometer on
Paranal is now in a condition to not only determine accurate {\it mean}
angular diameters of nearby Cepheid variables, but to follow their
angular diameter {\it variations} with high precision.
Using the Palomar Testbed Interferometer,
\citet{lane00,lane02} resolved the pulsation of the Cepheids $\zeta$~Gem and
$\eta$~Aql as early as 2000, but the comparison we present in this
letter is the first where error bars on the derived distance and
linear diameter are directly comparable at a few percent level between
the interferometric and IRSB techniques.

The star that we will discuss in this letter, $\ell$~Car, is the brightest Cepheid
in the sky. Its long period of about 35.5 days implies a large
mean diameter, which together with its relatively short
distance makes it an ideal target for resolving its angular diameter
variations with high accuracy. In this paper, we will compare
the interferometrically determined angular diameter curve of $\ell$~Car
with that determined from the IRSB technique,
and we will demonstrate that the two sets of angular diameters are
in excellent agreement. Based on the available high-precision angular
diameter and radial velocity curves for this star, we will also derive
a revised value of its distance and mean radius.
 
Several authors \citep{Sasselov94, Marengo03, Marengo04}
have pointed out potential sources of systematic uncertainties
in the determination of Cepheid distances using the interferometric
BW method. In particular, imperfections in the numerical
modeling of Cepheid atmospheres could lead to biased estimates
of the limb darkening and projection factor. We will discuss the
magnitude of these uncertainties in the case of $\ell$~Car.

\section{Interferometric observations}

The interferometric observations of $\ell$~Car were obtained with the
VLT Interferometer \citep{Glindemann00}, using its
commissioning instrument VINCI \citep{Kervella00,Kervella03} and 0.35\,m test siderostats.
This instrument recombines the light from two telescopes in the infrared
$K$ band (2.0--2.4\,$\mu$m), at an effective wavelength of $2.18\,\mu$m.
A detailed description of the interferometric data recorded on $\ell$~Car
can be found in K04.

The limb darkening (LD) models used to derive the photospheric diameters from
the fringe visibilities were taken from \citet{Claret00}.
The correction introduced on the uniform disk (UD) interferometric measurements
by the limb darkening is small in the $K$ band: for $\ell$~Car, we determine
$k= \theta_{\rm UD}/\theta_{\rm LD} = 0.966$. Considering the magnitude of this
correction, a total systematic uncertainty of $\pm 1\,\%$ appears reasonable.
However, until the limb darkening of a sample of Cepheids has been measured
directly by interferometry, this value relies exclusively on numerical models of the
atmosphere. This is expected to be achieved in the next years using for instance
the longest baselines of the VLTI (up to 202\,m) and the shorter $J$ and $H$ infrared
bands accessible with the AMBER instrument \citep{Petrov00}.

The LD correction is changing slightly over the pulsation of the star, due
to the change in effective temperature, but \citet{Marengo03} have estimated the
amplitude of this variation to less than $0.3\,\%$ peak to peak in the $H$ band
(for the 10\,days period Cepheid $\zeta$~Gem). It is even lower in the $K$ band,
and averages out in terms of rms dispersion. As a consequence,
we have neglected this variation in the present study.

The limb-darkened angular diameter measurements
are listed in Table~\ref{table_angdiams_l_car}.
Two error bars are given for each point, corresponding respectively
to the statistical uncertainty (internal error) and to the systematic
error introduced by the uncertainties on the assumed angular diameters
of the calibrator stars (external error). The phases given in
Table~\ref{table_angdiams_l_car} are based on the new ephemeris derived
in Sect.\ref{sec.diameter}.
These measurements were obtained during the commissioning of the VLTI,
and part of them are affected by relatively large uncertainties (3-5\%)
due to instrumental problems.  However, the precision reached by VINCI
and the test siderostats on this baseline is of the order of 1\%
on the angular diameter, as demonstrated
around the maximum diameter phase.

\section{The infrared surface brightness technique}

The IRSB technique has been presented and
discussed in detail in \citet{FG97} (hereafter FG97).
In brief, the angular diameter curve of a
given Cepheid variable is derived from its $V$ light and $(V-K)$ color
curve, appropriately corrected for extinction.
It is then combined with its linear displacement, which is essentially
the integral of the radial velocity curve. 
A linear regression of pairs of angular diameters and linear
displacements, obtained at the same pulsation phases, yields both the
distance, and the mean radius of the star.

While there are several sources of systematic uncertainty in the method,
as discussed in \citet{GFG97}, one of its great advantages is
its strong insensitivity to the adopted reddening corrections, and to
the metallicity of the Cepheid \citep{Storm04}.
With excellent observational data at
hand, individual Cepheid distances and radii can be determined with an
accuracy of the order of 5\,\% {\em if} the adopted $K$-band surface
brightness-color relation is correct.

A first calibration of this relation coming directly from interferometrically
determined angular diameters of Cepheid variables was presented by
\citet{Nordgren2002} (herafter N02). They found a satisfactory agreement
with the FG97 calibration, within the combined 1\,$\sigma$ uncertainties
of both surface brightness-color calibrations.
Considering more closely the results from N02,
an even better agreement is found between the $F_V(V-K)$
relations \emph{before} the zero point if forced between the different colors.
Before this operation, N02 found the relation:
\begin{equation}
F_{V_0} = 3.956 \pm 0.011 - 0.134 \pm 0.005 (V - K)_0
\end{equation}
that translates, after forcing the zero point to the average
of the three selected colors, to the relation:
\begin{equation}
F_{V_0} = 3.941 \pm 0.004 - 0.125 \pm 0.005 (V - K)_0.
\end{equation}
On the other hand, FG97 obtain:
\begin{equation}
F_{V_0} = 3.947 \pm 0.003 - 0.131 \pm 0.003 (V - K)_0
\end{equation}
From this comparison, it appears that the slope initially
determined by N02 for $F_V(V-K)$ is significantly different both from their
final value and from the FG97 relation. This difference could cause a
bias due to the averaging of the multi-color zero points.
Though small in absolute value, such a bias is of particular importance 
for $\ell$~Car, due to its relatively large $(V-K)$ color.

Another argument in favor of the FG97 surface brightness relation
is that it relies on a sample of 11 Cepheids with periods of 4 to 39~days,
while the relations established by N02 were derived from the observations
of only 3 Cepheids with periods of 5 to 10~days.
Such short period Cepheids are significantly hotter
than $\ell$~Car ($P = 35.5$~days), and a local difference of the
slope of the IRSB relations cannot be excluded.
For these two reasons, we choose to retain the FG97 calibration
for our analysis of the Cepheid $\ell$~Car in the following section.

\section{Diameter and distance}
\label{sec.diameter}

\subsection{Angular diameter}
\label{subsec.angular}

\begin{figure}[htp]
\plotone{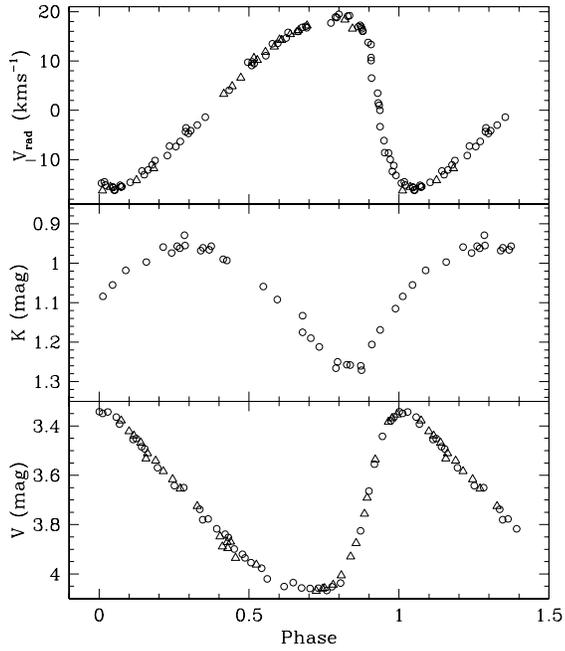}
\caption{\label{fig.VK}
Radial velocity curve of $\ell$~Car (upper panel)
using data from \citet{Taylor97} shifted by $-1.5$~\kms
(circles) and from \citet{Bersier02} (triangles).
The $K$ band photometric measurements (middle panel) were taken from
\citet{LS92}. We have relied on \citet{Pel76} (circles) and
\citet{Bersier02} (triangles) for the $V$ band data.}
\end{figure}

We have combined the photometric data from \citet{Pel76} and
\citet{Bersier02} to construct the $V$-band light curve for
$\ell$~Car. The two data sets are spanning almost 30 years
and allow an improved determination of the period of this variable. We
find $P=35.54804$~days. The time of maximum $V$ light has been adopted
from \citet{Szabados89} who give a value of $T_0=2440736.230$
which is also in good agreement with the more recent data. The
resulting light curve is shown in Fig.\ref{fig.VK}.
The $K$ band light curve is based on the data from
\citet{LS92} and is also shown in Fig.\ref{fig.VK}. The $(V-K)$
color curve which is needed by the IRSB method has been
constructed on the basis of the observed $V$ band data and a Fourier fit to
the $K$-band data as described in \citet{Storm04}.

For the radial-velocity curve we have used the data from
\citet{Taylor97} and \citet{Bersier02}.  Using the
new ephemeris from above we detected a slight offset of 1.5~\kms
between the two datasets. We choose to shift the
\citet{Taylor97} dataset by $-1.5$~\kms to bring all the data on
the well established CORAVEL system of \citet{Bersier02}.  We
note that the exact radial velocity zero point is irrelevant as the
method makes use of \emph{relative} velocities. The combined
radial velocity data are displayed in Fig.\ref{fig.VK}.

The application of the IRSB method has
followed the procedure described in \citet{Storm04}.
We have adopted the same reddening law with
$R_V=3.30$ and $R_K=0.30$, a reddening of $E(B-V)=0.17$ 
\citep{Fernie90}, and a projection factor, $p$, from radial to
pulsational velocity of $p=1.39-0.03\log P=1.343$ \citep{Hindsley86,GBM93}.
As discussed by \citet{Storm04} we only consider the
points in the phase interval from $0.0$ to $0.8$ (phase zero is
defined by the $V$ band maximum light).
We have applied a small phase shift of $-0.025$ to the radial velocity
data to bring the photometric and radial velocity based angular
diameters into agreement.  We note that a similar phase shift can be
achieved by lowering the systemic velocity by $1.5$~\kms.

The angular diameter curve obtained from the photometry has been
plotted in Fig.\ref{fig.svb}, together with the linear displacement
curve.  The photometric and interferometric diameter curves are
directly compared in Fig.~\ref{fig.LDfit}, where they are plotted as a
function of phase.
With these data we can compute the average
angular diameters obtained from each technique. 
For the IRSB, we find an average limb darkened angular diameter
$\overline{\theta_{\rm LD}} = 2.974 \pm 0.046$ milliarcsecond, and for
the interferometric measurements we find $\overline{\theta_{\rm LD}} =
2.992 \pm 0.012$ milliarcsecond. The
agreement between these two values is strikingly good.  This is a
serious indication that the calibration of the surface
brightness-color relation (FG97), based on non-pulsating giant stars,
does apply to Cepheids.

\subsection{Distance}
\label{subsec.distance}

The surface brightness method yields a
distance of $560 \pm 6$~pc, and a mean radius of $R=179 \pm 2$~\Rsolar.
The corresponding mean absolute $V$ magnitude is
$M_V=-5.57 \pm 0.02$~mag and the distance modulus is $(m-M)_0 = 8.74
\pm 0.05$.  The error estimates are all intrinsic 1$\sigma$ random
errors.  In addition to these random errors, a systematic error of the
order of 4\% should be taken into account, as discussed by \citet{GFG97}.
The final IRSB values are thus $d = 560 \pm 23$~pc,
and $R = 179 \pm 7$~\Rsolar.
Compared to \citet{Storm04} we find a significantly
(0.24~mag) shorter distance modulus for $\ell$~Car. This can be
explained by the use in the present Letter of the new and superior
radial velocity data from \citet{Taylor97} and \citet{Bersier02}.

K04 found $d = 603^{+24}_{-19}$ pc, using the interferometric angular diameters
and a subset of the radial velocity data used here. To make the
comparison more relevant, we determined the distance and radius using
the same data (interferometric diameters from
Table~\ref{table_angdiams_l_car} and radial velocity from
\citet{Taylor97} and \citet{Bersier02} -- see above),
the same ephemeris, and the same
projection factor (see Sect~\ref{subsec.angular}). Using the method of
K04, we find a distance $d = 566^{+24}_{-19}$, and a linear radius
$R = 182^{+8}_{-7}$~\Rsolar. This is in excellent agreement with the values
obtained from the IRSB method.

This 6\,\% difference in the distances based on interferometric diameters
(603\,pc for K04 versus 566\,pc here) has two major causes. First, the
$p$-factor used in the present paper is $\sim 1.3$\% smaller than in K04.
The choice of the reference used for the $p$-factor has currently an impact
of a few percents on its value. This indicates that the average value of the
$p$-factor for a given Cepheid is currently uncertain by at least a similar
amount, and this systematic error translates linearly to the distance
determination.

Secondly, the use of a different -- and superior -- data set for the
radial velocity makes the radius curve different from K04. In
particular the amplitude is smaller here than in K04 by $\sim 3$\%. This
is likely due to the more complete phase coverage that we have here, and
possibly also to a different choice of spectral lines to estimate
the radial velocity. This amplitude difference translates linearly on the
distance through the BW method.

\section{Conclusion}

\begin{figure}[htp]
\plotone{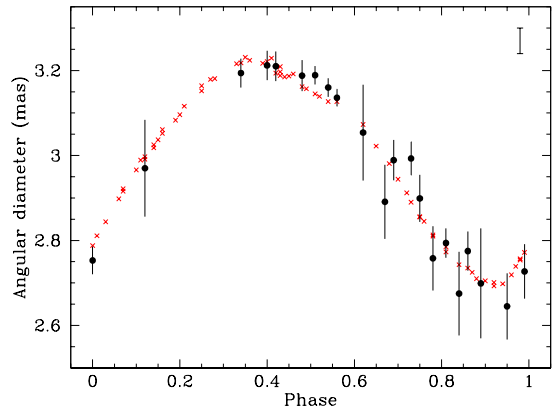}
\caption{\label{fig.LDfit} The interferometrically determined angular
diameters, plotted against phase (solid dots) with the angular
diameters derived with the infrared surface brightness method overplotted
(crosses). In the upper right corner a typical error bar 
for the surface brightness method data is shown.}
\end{figure}

The main point of our paper is to show that with a
consistent treatment of the data, the internal accuracy of both methods
(IRSB or interferometry) is extremely good: the angular diameter variation
observed using the VLTI agrees very well with that derived from the
$F_V(V-K)$ version of the IRSB technique as calibrated by FG97.
For all the interferometric measurements, the corresponding IRSB angular
diameter at the same phase lies within the combined 1$\sigma$ error
bars of the two measurements (Fig.~\ref{fig.LDfit}).
Even more importantly, the mean angular diameter of the Cepheid as
derived from both independent sets of angular diameter determination
are in excellent agreement, within a few percents.

Unfortunately, this is not equivalent to say that the Cepheid
distance scale is calibrated to a 1\% accuracy. We have drawn
attention to remaining sources of systematic errors that can
affect Cepheid radii and distances up to several percents.
As an illustration of these sources, K04 obtain a distance $d= 603^{+24}_{-19}$\,pc
for $\ell$~Car, while we obtain $d = 566^{+24}_{-19}$\,pc from the same
interferometric data.

\begin{figure}[htp]
\plotone{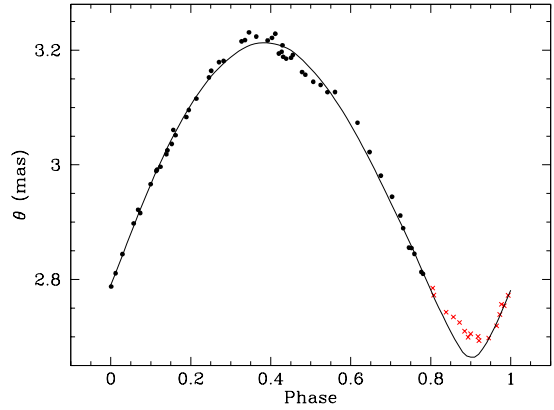}
\caption{\label{fig.svb} Photometric angular diameters plotted against
phase for our best fitting distance.  The solid curve represents the
integrated radial velocity curve of $\ell$~Car for the adopted
distance.}
\end{figure}

We have already shown that most of the 6\,\% difference
(equivalent to $1.3\,\sigma$) can be explained by the use of different radial
velocity data and projection factor. Another thing to consider is the phase interval
used. K04 used measurements over the whole
pulsation cycle whereas in the IRSB technique, one
avoids the phase interval 0.8--1 (Fig.~\ref{fig.svb}).
During that phase interval, that corresponds to the rebound of
the atmosphere around the minimum radius, energetic shock waves are
created. As discussed by \citet{Sabbey95}, they produce asymmetric line
profiles in the Cepheid spectrum. Recent modeling using a self-consistent dynamical
approach also show that the $\tau=1$ photosphere may not be comoving with
the atmosphere of the Cepheid during its pulsation, at the 1\,\% level \citep{Nardetto04}.
Such an effect would impact the $p$-factor, modify the shape of
the radial velocity curve, and thus bias the amplitude of the radius variation,
possibly up to a level of a few percents. As the BW method (either its classical or
interferometric versions) relies linearly on this amplitude, a bias at this level
presently cannot be excluded.

The interferometric BW method is currently limited to
distances of 1--2 \,kpc, due to the limited length of the available
baselines. The IRSB technique on the other hand 
can reach extra-galactic Cepheids as already demonstrated by
\citet[for the LMC]{gieren00} and by \citet[for the SMC]{Storm04}.
Using high precision interferometric measurements of $\ell$~Car and
other Cepheids, it will be possible to calibrate the IRSB
method down to the level of a few percents.
From the present comparison, we already see that
this fundamental calibration will be very similar to the calibration
found by FG97 and \citet{Nordgren2002}.

\acknowledgments

WPG acknowledges support for this work from the chilean FONDAP Center
for Astrophysics 15010003.

\begin{table}
\caption{Angular diameter measurements of $\ell$~Car. The statistical
and systematic calibration uncertainties are mentioned separately in brackets.
\label{table_angdiams_l_car}}
\begin{tabular}{ccc}
\tableline\tableline
\noalign{\smallskip}
Julian Date & Phase & $\theta_{\rm LD}$ (mas) \\
\tableline
\noalign{\smallskip}
 2452453.498 & 0.618 & $3.054 \pm 0.113\ _{[0.041, 0.105]}$\\
 2452739.564 & 0.665 & $2.891 \pm 0.087\ _{[0.076, 0.043]}$\\
 2452740.569 & 0.693 & $2.989 \pm 0.047\ _{[0.018, 0.044]}$\\
 2452741.717 & 0.726 & $2.993 \pm 0.039\ _{[0.026, 0.029]}$\\
 2452742.712 & 0.754 & $2.899 \pm 0.056\ _{[0.035, 0.043]}$\\
 2452743.698 & 0.781 & $2.758 \pm 0.076\ _{[0.074, 0.016]}$\\
 2452744.634 & 0.808 & $2.794 \pm 0.035\ _{[0.032, 0.013]}$\\
 2452745.629 & 0.836 & $2.675 \pm 0.098\ _{[0.097, 0.017]}$\\
 2452746.620 & 0.864 & $2.775 \pm 0.046\ _{[0.023, 0.040]}$\\
 2452747.599 & 0.891 & $2.699 \pm 0.129\ _{[0.127, 0.026]}$\\
 2452749.576 & 0.947 & $2.645 \pm 0.078\ _{[0.077, 0.012]}$\\
 2452751.579 & 0.003 & $2.753 \pm 0.033\ _{[0.028, 0.017]}$\\
 2452755.617 & 0.117 & $2.970 \pm 0.113\ _{[0.113, 0.013]}$\\
 2452763.555 & 0.340 & $3.194 \pm 0.034\ _{[0.009, 0.033]}$\\
 2452765.555 & 0.396 & $3.212 \pm 0.034\ _{[0.011, 0.033]}$\\
 2452766.550 & 0.424 & $3.210 \pm 0.035\ _{[0.011, 0.033]}$\\
 2452768.566 & 0.481 & $3.188 \pm 0.037\ _{[0.011, 0.035]}$\\
 2452769.575 & 0.509 & $3.189 \pm 0.022\ _{[0.018, 0.012]}$\\
 2452770.535 & 0.536 & $3.160 \pm 0.022\ _{[0.020, 0.009]}$\\
 2452771.528 & 0.564 & $3.136 \pm 0.020\ _{[0.017, 0.010]}$\\
 2452786.620 & 0.989 & $2.727 \pm 0.064\ _{[0.012, 0.063]}$\\
\noalign{\smallskip}
\tableline
\end{tabular}
\end{table}


\newpage

\end{document}